\documentstyle[prb,aps,floats,epsfig,amssymb,amsmath]{revtex}
\def\sech{{\rm sech}}
\def\atanh{{\rm atanh}}

\begin{document}
\twocolumn[\hsize\textwidth\columnwidth\hsize\csname@twocolumnfalse
\endcsname
\title{Introduction to the Bethe Ansatz II} 
\author{Michael Karbach$^*$, Kun Hu$^\dagger$, and Gerhard M\"uller$^\dagger$} 
\address{$^*$Bergische Universit\"at Wuppertal, Fachbereich Physik, 
         D-42097 Wuppertal, Germany \\ 
         $^\dagger$Department of Physics, University of Rhode Island,
         Kingston RI 02881-0817}
\date{\today}
\maketitle
\begin{abstract}
\end{abstract}
\twocolumn
]
\section{Introduction}
Quantum spin chains are physically realized in
quasi-one-dimensional (1D) magnetic insulators. In such materials,
the magnetic ions (for example, $\rm Cu^{++}$ and $\rm Co^{++}$
with effective spin $s=\frac{1}{2}$, $\rm Ni^{++}$ with $s=1$, and
$\rm Mn^{++}$ with $s=\frac{5}{2}$) are arranged in coupled linear
arrays which are well separated and magnetically screened from each
other by large non-magnetic molecules. Over the years,
magneto-chemists have refined the art of designing and growing
crystals of quasi-1D magnetic materials to the point where physical
realizations of many a theorist's pet model can now be custom-made.

Most prominent among such systems are realizations of the 1D
spin-1/2 Heisenberg antiferromagnet,
\begin{equation}\label{eq:H}
H_A = J\sum_{n=1}^N {\bf S}_n \cdot {\bf S}_{n+1},
\end{equation}
the model system which inspired Bethe to formulate the now
celebrated method for calculating eigenvalues, eigenvectors, and a
host of physical properties. Interest in quasi-1D magnetic
materials has stimulated theoretical work on quantum spin chains
from the early sixties until the present. Some of the advances
achieved via the Bethe ansatz emerged in direct response to
experimental data which had remained unexplained by standard
approximations used in many-body theory.

In Part I of this series\cite{KM97} we introduced the Bethe ansatz
in the spirit of Bethe's original 1931 paper.\cite{Beth31} The
focus there was on the Hamiltonian $H_F=-H_A$, where the negative
sign makes the exchange coupling ferromagnetic. The ferromagnetic
ground state has all the spins aligned and is
$(N+1)$-fold degenerate. The reduced rotational symmetry of every
ground state vector relative to that of the Hamiltonian reflects
the presence of ferromagnetic long-range order at zero temperature.
One vector of the ferromagnetic ground state, $|F\rangle \equiv
|\!\!\uparrow\!\cdots\!\uparrow\rangle$, in the notation of Part I,
serves as the reference state (vacuum) of the Bethe ansatz. All
other eigenstates are generated from $|F\rangle$ through multiple
magnon excitations.

We also investigated some low-lying excitations which involve only a
small number of magnons. For example, the Bethe ansatz enabled us
to study the properties of a branch of two-magnon bound-states and a
continuum of two-magnon scattering states, and gave us a perfect
tool for understanding the composite nature of these states in
relation to the elementary particles (magnons).

In this column we turn our attention to the Heisenberg
antiferromagnet $H_A$. All the eigenvectors remain the same as in
$H_F$, but the energy eigenvalues have the opposite sign.
Therefore, the physical properties are very different, and the state
$|F\rangle$ now has the highest energy. Our immediate goals are to
find the exact ground state $|A\rangle$ of $H_A$, to investigate
how the state
$|A\rangle$ gradually transforms to the state $|F\rangle$ in the
presence of a magnetic field of increasing strength, and to prepare
the ground work for a systematic study of the excitation spectrum
relative to the state $|A\rangle$.

As in Part I we will emphasize computational aspects of the Bethe
ansatz. The numerical study of finite systems via the Bethe ansatz
is akin to a simulation in many respects, and yields important
insights into the underlying physics.

\section{Ground state}

What is the nature of the ground state state $|A\rangle$? How do we
find it? Is its structure as simple as that of $|F\rangle$? An
obvious candidate for $|A\rangle$ is the N\'eel state. The two
vectors
\begin{equation}\label{eq:N}
|{\mathcal N}_1\rangle \equiv
|\!\uparrow\downarrow\uparrow\cdots\downarrow\rangle,~~
|{\mathcal N}_2\rangle \equiv
|\!\downarrow\uparrow\downarrow\cdots\uparrow\rangle
\end{equation}
reflect antiferromagnetic long-range order in its purest form just
as the vector $|F\rangle$ does for ferromagnetic long-range order.
Henceforth we assume that the number of spins $N$ in (\ref{eq:H}) is
even and that periodic boundary conditions are imposed.

Inspection shows that neither $|{\mathcal N}_1\rangle,|{\mathcal
N}_2\rangle$, nor the translationally invariant linear combinations,
$|{\mathcal N}_\pm\rangle = (|{\mathcal N}_1\rangle \pm |{\mathcal
N}_2\rangle)/\sqrt{2}$, are eigenvectors of $H_A$. In the energy
expectation value $\langle H_A\rangle$, the N\'eel state minimizes
$\langle S_n^zS_{n+1}^z\rangle$ but not $\langle
S_n^xS_{n+1}^x\rangle$ and $\langle S_n^yS_{n+1}^y\rangle$ 
(Problem~\ref{prb:1}). A state with the full rotational symmetry of
$H_A$ can have a lower energy. Like
$|{\mathcal N}_\pm\rangle$, the ground state $|A\rangle$ will be
found in the subspace with
$S_T^z \equiv \sum_nS_n^z = 0$. Because of their simplicity, the
N\'eel states are convenient starting vectors for the computation
of the finite-$N$ ground state energy and wave function via
standard iterative procedures (steepest-descent and
conjugate-gradient methods).\cite{NVM93} However, here we take a
different route.

In the framework of the Bethe ansatz, all eigenstates of $H_A$ with $S_T^z=0$
can be obtained from the reference state $|F\rangle$ by exciting $r\equiv N/2$
magnons with momenta $k_i$ and (negative) energies $-J(1-\cos k_i)$. The exact
prescription was stated in Part I. Each eigenstate is specified
by a different set of $N/2$ Bethe quantum numbers $\{\lambda_i\}$.
The momenta $k_i$ and the phase angles $\theta_{ij}$ in the
coefficients (I28) of the Bethe wave function (I27) result
from the Bethe ansatz equations (I33) and (I35).\cite{note1} A
finite-$N$ study indicates that the ground state
$|A\rangle$ has real momenta $k_i$ and Bethe quantum numbers
(Problem~\ref{prb:2})
\begin{equation}\label{eq:lambda}
\{\lambda_i\}_A = \{1, 3, 5, \ldots, N-1\}.
\end{equation}

In Part I we worked directly with $k_i$ and $\theta_{ij}$, which
represent physical properties of the elementary particles (magnons)
created from the vacuum $|F\rangle$. At the
opposite end of the spectrum, the state $|A\rangle$, generated from
$|F\rangle$ via multiple magnon excitations, will be configured
as a new physical vacuum for $H_A$. The entire spectrum of $H_A$
can then be explored through multiple excitations of a different
kind of elementary particle called the {\it spinon}.

Computational convenience suggests that we replace the two sets of variables
$\{k_i\}$ and $\{\theta_{ij}\}$ by a single set of (generally complex)
variables $\{z_i\}$. If we relate every magnon momentum $k_i$ to a new
variable $z_i$ by
\begin{equation}
\label{eq:zjk}
k_i\equiv \pi -\phi(z_i),
\end{equation}
via the function $\phi(z) \equiv 2\arctan z$, then the relation (I33) between
every phase angle $\theta_{ij}$ for a magnon pair and the difference $z_i-z_j$
involves $\phi(z)$ again:
\begin{equation}
\label{eq:theta-zij}
\theta_{ij} = \pi \, {\rm sgn}[{\Re}(z_i-z_j)]
              - \phi\bigl[(z_i-z_j)/2\bigr].
\end{equation}
Here ${\Re}(x)$ denotes the real part of $x$, and sgn$(y)=\pm 1$
denotes the sign of $y$. Relations (\ref{eq:zjk}) and
(\ref{eq:theta-zij}) substituted into (I35) yield the Bethe ansatz
equations for the variables $z_i$:~\cite{YY66}
\begin{equation}
\label{eq:bae}
N\phi(z_i) = 2\pi I_i +
\sum_{j\neq i}\phi\bigl [(z_i-z_j)/2\bigr ],
\quad i=1,\ldots,r.
\end{equation}
The new Bethe quantum numbers $I_i$ assume integer values for odd $r$ and
half-integer values for even $r$. The relation between the sets $\{\lambda_i\}$
and $\{I_i\}$ is subtle. It depends, via ${\rm sgn}[{\Re}(z_i~-~z_j)]$, on the
configuration of the solution $\{z_i\}$ in the complex plane. For the ground
state $|A\rangle$, we obtain (Problem~\ref{prb:3}a)
\begin{equation}\label{eq:bqn}
\{I_i\}_A =
\biggl\{-\frac{N}{4}+\frac{1}{2},~-\frac{N}{4}+\frac{3}{2},\ldots,
 \frac{N}{4}-\frac{1}{2}\biggr\}.
\end{equation}

Given the solution $\{z_1,\ldots,z_r\}$ of Eqs.~(\ref{eq:bae}) for a state
specified by $\{I_1,\ldots,I_r\}$, its energy and wave number are
(Problem~\ref{prb:3}b)
\begin{mathletters}
\begin{eqnarray}\label{eq:e}
(E-E_F)/J &=& \sum_{i=1}^{r}\varepsilon(z_{i}), \\
k = \sum_{i=1}^r\bigl[\pi - \phi(z_i)\bigr]
&=& \pi r - \frac{2\pi}{N}\sum_{i=1}^r I_i, \label{eq:k}
\end{eqnarray}
\end{mathletters}with $E_F=JN/4$. The quantity $\phi(z_i)$ is
called the magnon {\it bare momentum}, and $\varepsilon(z_{i}) = dk_i/dz_i =
-2/(1+z_i^{2})$ is the magnon {\it bare energy}. The Bethe wave function (I27)
is obtained from the $\{z_i\}$ via (\ref{eq:zjk}) and (\ref{eq:theta-zij}).

The ground state $|A\rangle$ belongs to a class of eigenstates which
are characterized by real solutions $\{z_i\}$ of the Bethe ansatz
equations. To find them numerically we convert Eqs.~(\ref{eq:bae})
into an iterative process:
\begin{equation}
\label{eq:bae-it-n}
  z_i^{(n+1)} = \tan\biggl( \frac{\pi}{N} I_i +
  \frac{1}{2 N} \sum_{j\neq i}
   \phi\bigl[(z_i^{(n)}-z_j^{(n)})/2\bigr] \biggr).
\end{equation}
Starting from
$z_i^{(0)}=0$, the first iteration yields
$z_i^{(1)} = \tan(\pi I_i/N)$. Convergence toward the roots of
(\ref{eq:bae}) is fast. High-precision solutions $\{z_i\}$ can be
obtained on a personal computer within seconds for systems with up
to $N=256$ sites and within minutes for much larger systems (Problem
\ref{prb:4}). The ground-state energy per site inferred from
(\ref{eq:e}) is listed in Table~I for several lattice sizes.

\begin{table}[htb]
\caption{Numerical results for the energy per site of the ground
state $|A\rangle$ relative to the energy $E_F/JN=\frac{1}{4}$ of the
state $|F\rangle$ for various values of $N$ obtained via
$n_{\rm max}$ iterations of (\ref{eq:bae-it-n}). The CPU time quoted
is for a Pentium 130 computer running GNU C on Linux. The exact
result for
$N\to\infty$ is
$-\ln 2$.}
\medskip
\begin{tabular}{|rrrc|}
$N$      & $(E_A-E_F)/JN$  & $n_{\rm max}$ & CPU-time [sec] \\
\hline\hline
  16     & -0.696393522538549  &  38 & 0.01 \\
  64     & -0.693348459146139  &  83 & 0.08 \\
 256     & -0.693159743366446  & 195 & 2.38 \\
1024     & -0.693147965376242  & 483 & 104  \\
4096     & -0.693147229600349  &1256 &4695  \\ \hline
$\infty$ & -0.693147180559945  & --  & --
\end{tabular}
\end{table}

\begin{figure}[htb]
\centerline{\hspace{40mm}\epsfig{file=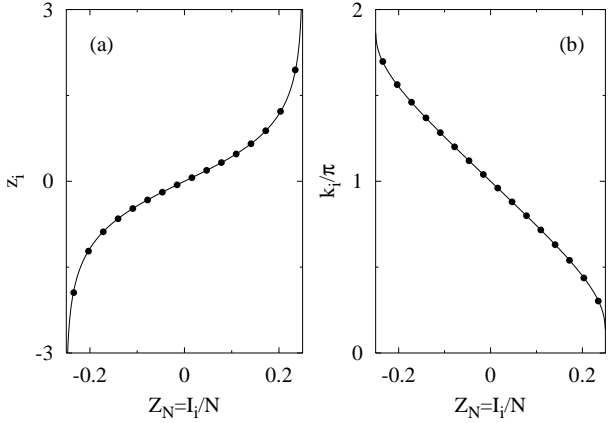,width=6.0cm,angle=-90}}
\caption{Ground state $|A\rangle$. (a) Solutions of
Eq.~(\ref{eq:bae}) and (b) magnon momenta in Eq.~(\ref{eq:zjk}) plotted
versus the rescaled Bethe quantum numbers (\ref{eq:bqn}) for
$N=32$ (circles) and $N=2048$ (lines).}
\label{fig:zI}
\end{figure}

In preparation of the analytical calculation which produces
$(E_A-E_F)/JN$ for $N\to\infty$, we inspect the $z_i$-configuration
for the finite-$N$ ground state
$|A\rangle$ obtained numerically. All roots are real, and their
values are sorted in order of the associated Bethe quantum numbers
$I_i$. The 16 circles in Fig.~\ref{fig:zI}(a) show $z_i$ plotted
versus $I_i/N$ for $N=32$. The solid line connects the
corresponding data for $N=2048$. The line-up of the finite-$N$ data
along a smooth monotonic curve suggests that the solutions of
(\ref{eq:bae}) can be described by a continuous distribution of
roots for
$N\to\infty$.

We give the inverse of the discrete function shown in Fig.~\ref{fig:zI}(a) the
name $Z_N(z_i) \equiv I_i/N$ and rewrite Eqs.~(\ref{eq:bae}) in the form:
\begin{equation} \label{eq:ZN}
2\pi Z_N(z_i) = \phi( z_i) -
\frac{1}{N} \sum_{j\neq i}\phi\bigl[(z_i-z_j)/2\bigr].
\end{equation}
For $N\to\infty$, $Z_N(z_i)$ becomes a continuous function $Z(z)$ whose
derivative, $\sigma_0(z) \equiv dZ/dz$, represents the distribution of
roots. In
Eq.~(\ref{eq:ZN}) the sum $(1/N)\sum_j\ldots$ is replaced by the
integral
$\int \! dz'\sigma_0(z')\ldots$ Upon differentiation the continuous
version of (\ref{eq:ZN}) becomes the linear integral equation
\begin{equation}\label{eq:sigma-integral}
2\pi\sigma_0(z) = -\varepsilon(z) - (K\ast \sigma_0)(z)
\end{equation}
with the kernel $K(z)=4/(4+z^{2})$; $(K\ast \sigma_0)(z)$ is
shorthand for the convolution
$\int_{-\infty}^{+\infty}dz'K(z-z')\sigma_0(z')$. Fourier
transforming Eq.~(\ref{eq:sigma-integral}) yields an algebraic
equation for
$\tilde{\sigma}_0(u) \equiv \int_{-\infty}^\infty dz\, e^{iuz}
\sigma_0(z)$. Applying the inverse Fourier transform to its
solution yields the result
\begin{equation}
\label{eq:sigma-z}
\sigma_0(z) = \frac{1}{4} \sech(\pi z/2).
\end{equation}
The (asymptotic) ground-state energy per site as inferred from
(\ref{eq:e}),
\begin{equation}\label{eq:EA}
\frac{E_A-E_F}{JN} = \!
\int_{-\infty}^{+\infty} \! dz \, \varepsilon(z)\, \sigma_0(z) =
-\ln 2,
\end{equation}
is significantly lower than the energy expectation
value of the N\'eel states (Problem~\ref{prb:1}). The state $|A\rangle$ has
total spin $S_T=0$ (singlet). Unlike $|F\rangle$, it retains the full
rotational
symmetry of (\ref{eq:H}) and does not exhibit magnetic long-range
order.\cite{Baxt73} The wave number of $|A\rangle$, obtained from (\ref{eq:k}),
is $k_A=0$ for even $N/2$ and $k_A=\pi$ for odd $N/2$. The important result
(\ref{eq:EA}) was derived by Hulth\'en\cite{Hult38} in the early years of the
Bethe ansatz.

In Fig.~\ref{fig:zI}(b) we plot the magnon momenta $k_i$ of $|A\rangle$ as
inferred from (\ref{eq:zjk}) versus $I_i/N$ for the same data as used in
Fig.~\ref{fig:zI}(a). The smoothness of the curve reflects the fact that the
state $|A\rangle$ for $N\to\infty$ can be described by a continuous
$k_i$-distribution (Problem \ref{prb:5})
\begin{equation}\label{eq:rhok}
\rho_0(k) = \left[8\sin^2\frac{k}{2}\cosh
\left(\frac{\pi}{2}\cot\frac{k}{2}\right)\right]^{-1}.
\end{equation}

\section{Magnetic field}

In the presence of a magnetic field $h$, the Hamiltonian (\ref{eq:H}) must be
supplemented by a Zeeman energy:
\begin{equation}\label{eq:Hh}
H = H_A - hS_T^z.
\end{equation}
The two parts of $H$ are in competition. Spin alignment in the positive
$z$-direction is energetically favored by the Zeeman term, but any aligned
nearest-neighbor pair costs exchange energy. Given that $[H_A,S_T^z]=0$, the
eigenvectors are independent of $h$. The $2S_T+1$ components (with $|S_T^z|\leq
S_T$) of any $S_T$-multiplet fan out symmetrically about the original level
position and depend linearly on $h$.

The largest downward energy shift in each multiplet is experienced by the state
with $S_T^z=S_T$, and that shift is proportional to $S_T$. The state
$|A\rangle$, which has $S_T=0$, does not move at all, whereas the state
$|F\rangle$ with $S_T=N/2$ descends more rapidly than any other state. Even
though $|F\rangle$ starts out at the top of the spectrum, it is
certain to become the ground state in a sufficiently strong field.
The saturation field $h_S$ marks the value of $h$ where
$|F\rangle$ overtakes its closest competitor in the race of levels
down the energy axis.

The pattern in which levels with increasing $S_T^z$ become the ground state of
$H$ as $h$ increases depends on their relative starting position along the
energy axis. From the zero-field energies of this set of states, we will now
determine the magnetization $m_z \equiv S_{T}^{z}/N$ of the ground state as a
function of $h$.\cite{Grif64}

\begin{figure}[htb]
\hspace*{1.8cm}
\centerline{\epsfig{file=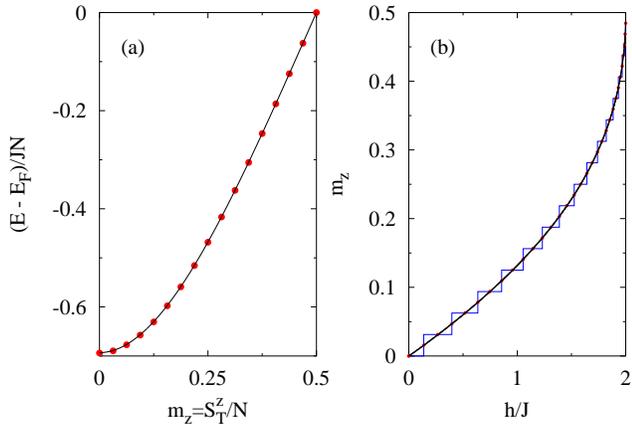,width=6.0cm,angle=-90}}
\caption{(a) Energy of the lowest states with a given $S_T^z$
at $h=0$. (b) Magnetization in the ground state of $H$.}
\label{fig:mag}
\end{figure}

The Bethe quantum numbers of the lowest state with quantum number $S_T^z=N/2-r
\geq 0$ are\cite{YY66}
\begin{equation}\label{eq:bqnh}
I_i = \frac{1}{2}\biggl(S_T^z - 1 + 2i -\frac{N}{2}\biggr),~~
i=1,\ldots,r,
\end{equation}
as can be confirmed by finite-$N$ studies of all states in the invariant
subspaces with $r=1,\ldots,N/2$. Using the iterative process
(\ref{eq:bae-it-n}), we can determine the energies of these states with high
precision (Problem~\ref{prb:6}). The red circles in Fig.~\ref{fig:mag}(a)
represent the quantity $[E(S_T^z) - E_F]/JN$ for $N=32$.  The solid line
connects the corresponding results for $N=2048$. For the finite-$N$ analysis it
is important to note that both the level energies $E(S_T^z)$ and the level
spacings $E(S_T^z)-E(S_T^z-1)$ increase monotonically with $S_T^z$.

At $h\neq 0$, all of these levels experience a downward shift of magnitude
$hS_T^z$. All spacings between adjacent levels shrink by the same amount $h$.
The first level crossing occurs between the state $|A\rangle$ with $S_T^z=0$
and
the state with $S_T^z=1$, which thereby becomes the new ground state. Next,
this
state is overtaken and replaced as the ground state by the state with $S_T^z=2$
and so forth. The last of exactly $N/2$ replacements involves the state with
$S_T^z=N/2-1$ and the state $|F\rangle$ with $S_T^z=N/2$. Their energy
difference in zero field is $2J$ independent of $N$ (see Part I). Consequently,
the saturation field is $h_S=2J$.

The magnetization $m_z$ grows in $N/2$ steps of width $1/N$ between $h=0$ and
$h=h_S$. In Fig.~\ref{fig:mag}(b) we plot $m_z$ versus $h$ for two system sizes
based on data obtained numerically. The blue staircase represents
the results for $N=32$. For $N=2048$ the step size has shrunk to
well within the thickness of the smooth curve shown.

An important observation is that the midpoints of the vertical and horizontal
steps of the $m_z(h)$ staircase (red dots) lie very close to the limiting
curve.
This behavior made it possible to extract quite accurate magnetization curves
for various spin chain models from very limited data.\cite{BF64}

Different scenarios are conceivable. For example, if the levels were arranged
in
the same sequence as in Fig.~\ref{fig:mag}(a) but with spacings increasing from
top to bottom, then the state with $S_T^z=0$ would be replaced directly by the
state with $S_T^z=N/2$.

In the thermodynamic limit, the energy per site of the lowest level with given
$S_T^z$ becomes the internal energy density at zero temperature,
\begin{equation}\label{eq:int-en}
u(m_z) = \lim_{N\to\infty}\frac{E(S_T^z)-E_F}{JN} .
\end{equation}
{From} (\ref{eq:int-en}) we obtain, via the thermodynamic relations,
\begin{equation}\label{eq:mzh}
h = \frac{du}{dm_z},~~
\chi_{zz} = \frac{dm_z}{dh} =
\biggl(\frac{d^2u}{dm_z^2}\biggr)^{\!-1}.
\end{equation}
The function $m_z(h)$ shown in Fig.~\ref{fig:mag}(b) is the inverse of the
derivative of the function $u(m_z)$ plotted in Fig.~\ref{fig:mag}(a). The slope
of $m_z(h)$ represents the longitudinal susceptibility $\chi_{zz}$. In a finite
system, where $m_z=S_T^z/N$ varies in steps of size $1/N$, Eqs.~(\ref{eq:mzh})
are replaced by
\begin{mathletters}
\begin{eqnarray}\label{eq:hmz}
h(m_z) &=& E(S_T^z) - E(S_T^z-1), \\
\label{eq:chizzE}
\chi_{zz}(m_z) &=&
\frac{1/N}{E(S_{T}^{z}+1)-2E(S_{T}^{z})+E(S_{T}^{z}-1)}.
\end{eqnarray}
\end{mathletters}

The data in Fig.~\ref{fig:mag}(b) indicate that $\chi_{zz}(h)$
has a nonzero value at $h=0$, grows monotonically with $h$, and
finally diverges at the saturation field $h=h_S$. The initial
value,\cite{Grif64,YY66}
\begin{equation}\label{eq:chi0}
\chi_{zz}(0) = \frac{1}{\pi^2J},
\end{equation}
turns out to be elusive to a straightforward slope analysis because of a
logarithmic singularity which produces an infinite curvature in $m_z(h)$ at
$h=0$ (Problem \ref{prb:7}). The divergence of
$\chi_{zz}(h)$ at
$h_S$ is of the type (Problem \ref{prb:8})
\begin{equation}\label{eq:chi-sat}
  \chi_{zz}(h) \stackrel{h\to h_{S}}{\longrightarrow}
  \frac{1}{2\pi}\frac{1}{\sqrt{J(h_{S}-h)}}.
\end{equation}
The characteristic upwardly bent magnetization curve with infinite
slope at the saturation field is a quantum effect unreproducible by
any simple and meaningful classical model system. The Hamiltonian
(\ref{eq:H}), reinterpreted as the energy function for coupled
three-component vectors, predicts a function $m_z(h)$ which
increases linearly from zero all the way to the saturation field.

\section{Two$-$spinon excitations}

\begin{figure}[htb]
\centerline{\epsfig{file=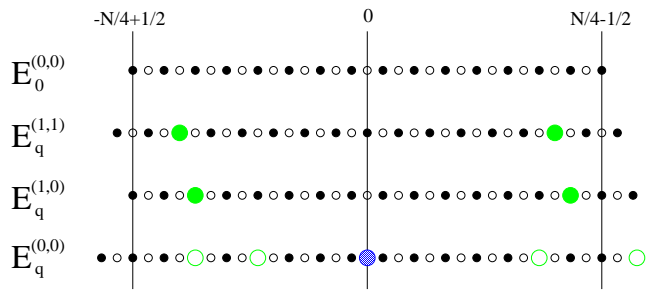,width=8.5cm,angle=0}}
\vspace*{0.4cm}
  \caption{Configurations of Bethe quantum numbers $I_i$ for the
    $N=32$ ground state (top row) and for one representative of three sets of
    two-spinon excitations with energy $E_q^{(S_T,S_T^z)}$. Each gap in the
    $I_i$-configurations of rows two and three (green full circles) represents a
    spinon. Each gap in row four (green open circles) represents half a spinon.
    The blue circle represents the Bethe quantum number associated with a pair
    of complex conjugate solutions, whereas all black circles are associated
    with real solutions.}
  \label{fig:Ii}
\end{figure}

Returning to zero magnetic field, let us explore the spectrum of
the low-lying excitations. From here on, the ground state
$|A\rangle$ (with $S_T^z=S_T=0$) replaces $|F\rangle$ (with
$S_T^z=S_T=N/2$) as the new reference state for all excited
states. The Bethe quantum numbers (\ref{eq:bqn}), which
characterize
$|A\rangle$, describe a perfectly regular array on the $I$-axis as illustrated
in the first row of Fig.~\ref{fig:Ii}. This array will be interpreted as a
physical vacuum. The spectrum of $H_A$ can then be generated systematically in
terms of the fundamental excitations characterized by elementary
modifications of this vacuum array.

In the subspace with $S_T^z=1$, a two-parameter set of states is obtained by
removing one magnon from the state $|A\rangle$. In doing so we eliminate one of
the $N/2$ Bethe quantum numbers from the set in the first row of
Fig.~\ref{fig:Ii} and rearrange the remaining $I_i$ in all configurations over
the expanded range $-\frac{1}{4}N \leq I_i\leq \frac{1}{4}N$.
Changing $S_T^z$ by one means that the $I_i$ switch from
half-integers to integers or vice versa. The number of distinct
configurations with $I_{i+1} - I_i \geq 1$ is
$\frac{1}{8}N(N+2)$. A generic configuration consists of three
clusters with two gaps between them as shown in the second row of
Fig.~\ref{fig:Ii}.

The two gaps in the otherwise uniform distribution of Bethe quantum numbers can
be interpreted as elementary particles (spinons) excited from the new physical
vacuum. The position of the gaps between the $I_i$-clusters determine the
momenta $\bar k_1, \bar k_2$ of the two spinons, which, in turn, add up to the
wave number of the two-spinon state relative to the wave number of
the vacuum:
$q\equiv k-k_A = \bar k_1+ \bar k_2$.

\begin{figure}[htb]
\vspace*{-1.2cm}
\centerline{\epsfig{file=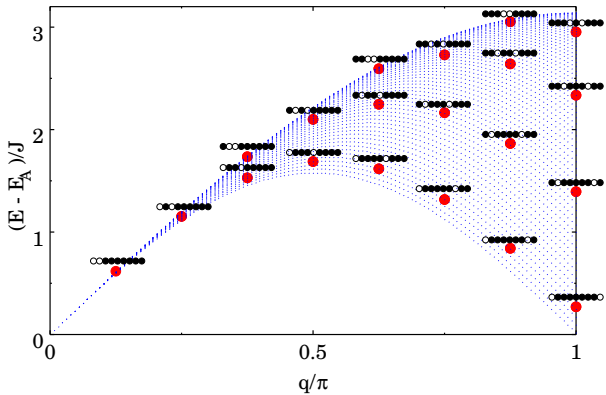,width=8cm,angle=0}}
\vspace*{1.7cm}
\caption{Two-spinon (triplet) excitations with $S_T^z=S_T=1$ for
  $N=16$ (red circles) and $N=256$ (blue dots). The configurations of (integer)
  Bethe quantum numbers $(-4 \leq I_1< \ldots < I_{7} \leq 4)$ for $N=16$ are
  symbolized by the black circles.}
\label{fig:2sp}
\end{figure}

What remains to be done for the finite-$N$ analysis is straightforward. We solve
the Bethe ansatz equations via (\ref{eq:bae-it-n}) for all the
$I_i$-configurations just specified. A plot of the two-spinon energies $E-E_A$
versus wave number $k-k_A$ for $N=16$ as inferred from the solutions $\{z_i\}$
via (\ref{eq:e}) and (\ref{eq:k}) is shown in Fig.~\ref{fig:2sp} (red circles).
Also shown are the Bethe quantum numbers for each excitation. The pattern is
readily recognized and extended (via reflection $I_i \leftrightarrow -I_i$) to
the other half of the Brillouin zone. We have drawn (in blue) the corresponding
two-spinon excitations for $N=256$. The $\frac{1}{4}N(\frac{1}{4}N+1)=4160$ dots
in the range $0<q\leq\pi$ produce a density plot for the two-spinon continuum
which emerges in the limit $N\to\infty$.

Compare this set of two-spinon scattering states with the set of
two-magnon states plotted in Fig.~2 of Part~I. There we found that pairs of
magnons form a continuum of two-magnon scattering states and a
branch of two-magnon bound states. Two-spinon bound states exist as
well and will be discussed in a later column.

A derivation of the exact lower and upper boundaries of the
two-spinon continuum,\cite{dCPYam}
\begin{equation}\label{eq:2spb}
\epsilon_L(q) = \frac{\pi}{2}J|\sin q|,~~
\epsilon_U(q) = \pi J|\sin\frac{q}{2}|,
\end{equation}
starts from the Bethe ansatz equations (\ref{eq:ZN}). We set $r=N/2-1$ and use
the Bethe quantum numbers from the second row of Fig.~\ref{fig:Ii}. When we
replace the sum by an integral, we must account for the gaps between the
$I_i$-clusters by two $1/N$-corrections. The result (after differentiation) is
an integral equation which differs from (\ref{eq:sigma-integral}) by two extra
terms related to the $I_i$-gaps:
\begin{eqnarray}\label{eq:2sp-inteq}
2\pi\sigma(z) &=& -\varepsilon(z) - (K\ast\sigma)(z) +
\frac{1}{N}\sum_{l=1,2} K(z-\bar z_{l}).
\end{eqnarray}
If we write $\sigma(z) = \sigma_0(z) + \sigma_1(z) + \sigma_2(z)$, where
$\sigma_0(z)$ is the solution (\ref{eq:sigma-z}) of
Eq.~(\ref{eq:sigma-integral}), the two-spinon corrections are
solutions of
\begin{equation}\label{eq:2sp-inteql}
2\pi\sigma_l(z) = \frac{1}{N}K(z-\bar
z_l)-(K\ast\sigma_{l})(z),~~l=1,2.
\end{equation}
Because Eqs.~(\ref{eq:2sp-inteql}) have the same integral kernel $K$ as
Eq.~(\ref{eq:sigma-integral}), the solutions of all three equations can be
expressed by the same resolvent $R$,
\begin{equation}\label{eq:012}
2\pi\sigma_l(z) = g_l(z) - (R\ast g_l)(z),~~l=0,1,2
\end{equation}
where $g_0(z)$ and $g_1(z), g_2(z)$ are the inhomogeneities of
Eqs.~(\ref{eq:sigma-integral}) and (\ref{eq:2sp-inteql}),
respectively. The resolvent, which does not depend on the
inhomogeneity, satisfies
\begin{equation}\label{eq:KR}
2\pi R(z) = K(z) - (K\ast R)(z).
\end{equation}
Because $g_l(z) = (1/N)K(z-\bar z_l)$ for $l=1,2$ makes (\ref{eq:2sp-inteql})
equivalent to (\ref{eq:KR}), we can write $\sigma_l(z) = (1/N)R(z-\bar z_l)$.
This is all we need to know about the solutions.

Using Eq.~(\ref{eq:e}) for the two-spinon energies, we must again
correct for the two $I_i$-gaps when we convert the sum into an integral:
\begin{equation}\label{eq:2spef}
E-E_F = NJ \! \int_{-\infty}^{+\infty} \!
dz\, \varepsilon(z)\sigma(z)
 - J \! \sum_{l=1,2} \varepsilon(\bar z_{l}).
\end{equation}
Subtracting $E_A-E_F$ yields
\begin{eqnarray}\label{eq:2spea}
E-E_A &=&
-J\sum_{l=1,2}[\varepsilon(\bar z_{l}) -
N(\sigma_{l}\ast\varepsilon)(0)]
 \nonumber \\
&=& J\sum_{l=1,2}\left[g_0(\bar z_l) -
(R\ast g_{0})(\bar z_{l})\right] \nonumber \\
&=& J\sum_{l=1,2} \sigma_0(\bar z_l)
= \frac{\pi J}{2}\sum_{l=1,2}\sech\left(\frac{\pi \bar
z_l}{2}\right).
\end{eqnarray}
Now we must relate the $I_i$-gaps, that is, the values $\bar z_1,
\bar z_2$ to the spinon momenta $\bar k_1, \bar k_2$. Starting from
the configuration in the first row of Fig.~\ref{fig:Ii}, we remove
one Bethe quantum number at or near the center. The accompanying
integer $\leftrightarrow$ half-integer switch of the remaining
$I_i$ opens a double gap at or near the center. Shifting several
$I_i$ from the left and right toward the center then produces the generic three
clusters. This change in configuration applied to (\ref{eq:k}) yields the
following two-spinon wave number relative to the vacuum:
\begin{eqnarray}\label{eq:2spka}
q &=& k-k_A = \bar k_1+ \bar k_2 = 
 \pi - \frac{2\pi}{N}\sum_{i=1}^{N/2-1}\Delta I_i \nonumber \\
&=& \sum_{l=1,2}\left[\frac{\pi}{2} - 2\pi\int_0^{\bar z_l}dz\,\sigma_0(z)\right]
\nonumber \\
&=& \sum_l\left[\frac{\pi}{2} - \arctan\left(\sinh\left(\frac{\pi
        \bar z_l}{2}\right)\right)\right],
\end{eqnarray}
or $\sin \bar k_l=\sech(\pi\bar z_l/2),\;l\!=\!1,2$. These relations
substituted into Eq.~(\ref{eq:2spea}) produce the two-spinon
spectrum
\begin{eqnarray}\label{eq:2speaqp}
E-E_A &=& \frac{\pi}{2}J[|\sin \bar k_1| + |\sin \bar k_2|] \nonumber \\
&=& \pi J|\sin(q/2)\cos(p/2)|
\end{eqnarray}
for $0\leq p\leq q$ and $0\leq q \leq 2\pi-p$. The spectrum is a two-parameter
continuum with boundaries (\ref{eq:2spb}).

Like magnons, spinons carry a spin in addition to energy and
momentum.\cite{FT81} Unlike magnons, which are spin-1 particles
(see Part I), spinons are spin-$\frac{1}{2}$ particles. In a chain
with even $N$, where all eigenstates have integer-valued $S_T^z$,
spinons occur only in pairs. The spins
$s_l=\pm\frac{1}{2}, l=1,2$ of the two spinons in a two-spinon
eigenstate of $H_A$ can be combined in four different ways to form
a triplet state $(S_T=1,\; S_T^z=0,\pm1)$ or a singlet state
$(S_T=S_T^z=0)$. They are described by distinct configurations of
Bethe quantum numbers.

We have already analyzed the spectrum of the two-spinon triplet
states with
$s_1=s_2=+\frac{1}{2}$ $(S_T=1,\; S_T^z=+1)$, as specified by
$I_i$-configurations of the kind shown in the second row of Fig.~\ref{fig:Ii}.
The two-spinon states with $s_1 = s_2 =-1$ $(S_T=1,\; S_T^z=-1)$
are obtained from these states by a simple spin-flip
transformation. They have exactly the same energy-momentum
relations.

The remaining two sets of two-spinon states have $s_1 = -s_2$, that
is,
$S_T^z=0$. One of these sets contains triplets $(S_T=1)$ and the
other singlets $(S_T=0)$. The former is obtained if we shift all
Bethe quantum numbers in the second row one position to the right
$(I_i \to I_i+\frac{1}{2},\; i=1,\ldots,r-1)$ as shown in the third
row and add one non-movable Bethe quantum number $I_r =
\frac{1}{4}N +
\frac{1}{2}$ to the set (Problem~\ref{prb:9}). The number of distinct
configurations is then the same as in row two and the corresponding states have
the same wave number and energy. Symmetry requires that all three components
$(S_T^z=0,\pm1)$ of the triplet states are degenerate.

Whereas all two-spinon triplets are described by real solutions
$\{z_i\}$, the two-spinon singlet states $(S_T^z=S_T=0)$ are
characterized by one pair of complex conjugate solutions $z_1
\equiv u+iv$, $z_2 \equiv u-iv$ in addition to the real solutions
$z_3,\ldots,z_{N/2}$.\cite{Woyn82b} Finding the
$I_i$-configurations for a particular set of eigenstates with
complex solutions (such as indicated in row four of
Fig.~\ref{fig:Ii}) and then solving the associated Bethe ansatz
equations turns out to be a much more delicate task than was the
identification and the calculation of purely real solutions.

A straightforward adaptation of the iterative process (\ref{eq:bae-it-n}) to
complex $z_i$ will, in general, not converge toward a physically meaningful
solution. Making the Bethe ansatz equations amenable to root-finding algorithms
which are adequate for this task requires that we convert Eqs.~(\ref{eq:bae})
into a set of equations with real solutions. For the two-spinon
singlets we arrive at the following set of $N/2$ equations for
$\{u,v,z_3,\ldots,z_{N/2}\}$ (Problem \ref{prb:10}):
\begin{mathletters}\label{eq:bae-cc-gen}
\begin{eqnarray}
\label{eq:bae-cc-1} 
N\phi(z_i) &=& 2\pi I_i^{(1)} + 
\sum_{\stackrel{\scriptstyle j=3}{j\neq i}}^{N/2} 
\phi\left(\frac{z_i-z_j}{2}\right)
 \nonumber \\ && \hspace*{-3cm} +
\phi\left(\frac{4(z_i-u)}{4-(z_{i}-u)^{2}-v^{2}}\right),
\quad i=3,\ldots,\frac{N}{2}, 
\\ \label{eq:bae-cc-2} 
 N\phi\left(\frac{2u}{1-u^{2}-v^{2}}\right) &=& 2\pi\left(I^{(2)} + N/2\right)
\nonumber \\ && \hspace*{-1cm}+ 
\sum_{i=3}^{N/2}\phi\left(\frac{4(u-z_i)}{4-(u-z_{i})^{2}-v^{2}}\right),
 \\ \label{eq:bae-cc-3b} 
N\varphi\left(\frac{2v}{1+u^{2}+v^{2}}\right) &=&
\varphi\left(\frac{2v}{1+v^2}\right) 
\nonumber \\ &&\hspace*{-1cm} +
\sum_{i=3}^{N/2}\varphi\left(\frac{4v}{4+v^{2}+(u-z_{i})^{2}}\right),
\end{eqnarray}
\end{mathletters}
where $\varphi(x) \equiv 2\, \atanh(x)$. Here we have replaced the first two
Bethe quantum numbers $I_1,I_2$ of Eq.~(\ref{eq:bae}) by a single (integer)
Bethe quantum number $I^{(2)}$ associated with $z_1=z_2^*$. The Bethe quantum
numbers associated with real solutions, now renamed $I_i^{(1)}$, are constrained
to the range $(-\frac{1}{4}N -\frac{1}{2} \leq I_i^{(1)} \leq \frac{1}{4}N
+\frac{1}{2})$ as indicated in the fourth row of Fig.~\ref{fig:Ii}. This
notation is used in preparation of a general classification of Bethe ansatz
solutions (string hypothesis) which is valid for $N\to\infty$ and will be
discussed in a later column.

The dependence of the wave number on the Bethe quantum numbers for all states
with one pair of complex solutions is then
\begin{equation}\label{eq:kcc}
 k = \pi(r-1) - \frac{2\pi}{N}I^{(2)} - \frac{2\pi}{N}\sum_{i=3}^r I_i^{(1)},
\end{equation}
where $r=N/2$ for the two-spinon singlets considered here.

\begin{figure}[htb]
\vspace*{0.2cm}
\centerline{\epsfig{file=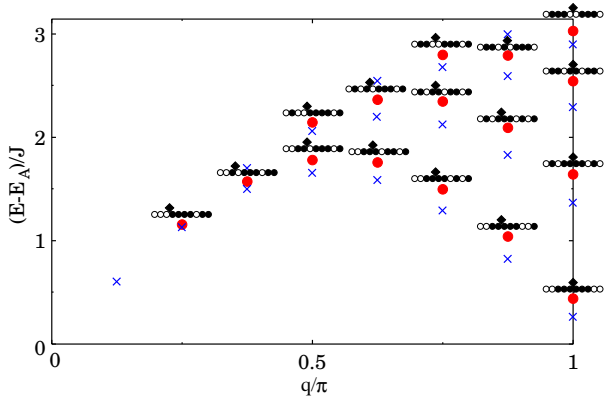,width=8cm,angle=0}}
\vspace*{0.4cm}
  \caption{Two-spinon (singlet) excitations with $S_T^z=S_T=0$ for
    $N=16$ (red circles). The configurations of (half-integer) Bethe quantum
    numbers $(-\frac{9}{2} \leq I_3 < \ldots < I_{8} \leq \frac{9}{2})$ are
    symbolized by the black circles and the Bethe quantum number $I^{(2)}$
    pertaining to the pair of complex solutions by black diamonds. The $N=16$
    triplets from Fig.~\ref{fig:2sp} are shown as blue crosses for comparison.}
\label{fig:2sps}
\end{figure}

Figure~\ref{fig:2sps} depicts the energy (\ref{eq:e}) versus the wave number
(\ref{eq:kcc}) of the two-spinon singlets (red circles) for $N=16$ and $0 < q
\leq \pi$. We use the same representation as in Fig.~\ref{fig:2sp} and show the
two-spinon triplets again for comparison. The number of two-spinon singlets
identified here is smaller, namely $\frac{1}{8}N(N-2)$ over the entire Brillouin
zone, than the number $\frac{1}{8}N(N+2)$ of two-spinon triplets identified
previously. Although the $I_i$-configurations of the singlets are more
complicated than those of the triplets, the pattern is still recognizable to the
extent that it may serve as a useful guide for applications to other system
sizes. The reader will find it easy to identify in the tables of Part I one
state for $N=4$ and three states for $N=6$ which belong to this class of singlet
excitations. In Table II we present the two-spinon singlet data for $N=8$.

\begin{table}[tb]\label{tab:2sps8}
\caption{Two-spinon singlet solutions of Eqs.~(\ref{eq:bae-cc-gen})
for $N=8$ and $0<q\leq\pi$.}
\medskip
\begin{tabular}{l|ll}
$I^{(2)}=-1$ & $q=\pi/2 $ & \hspace*{-1.5cm}$(E-E_{A})/J =
2.0330594201$ \\
$I_3^{(1)}=\frac{1}{2}$ & $u=-0.7040828497$ & $v=-1.0011163392$ \\
$I_4^{(1)}=\frac{5}{2}$ & $z_3=+0.2940252223$ & $z_4=+1.1141404771$ \\ \hline
$I^{(2)}=0$ & $q=3\pi/4 $ & \hspace*{-1.5cm}$(E-E_{A})/J
=1.9439866277 $ \\
$I_3^{(1)}=-\frac{1}{2}$ & $u=-0.4504249992$ & $v=-1.0000577243$ \\
$I_4^{(1)}=\frac{3}{2}$ & $z_3=-0.2423723559$ & $z_4=+1.1432223543$ \\ \hline
$I^{(2)}=0$ & $q=\pi $ & \hspace*{-1.5cm}$(E-E_{A})/J =
0.9514652606 $ \\
$I_3^{(1)}=-\frac{1}{2}$ & $u=0$ & $v=-1$ \\
$I_4^{(1)}=\frac{1}{2}$ & $z_3=-0.2849381356$ & $z_4=+0.2849381356$ \\ \hline
$I^{(2)}=0$ & $q=\pi $ & \hspace*{-1.5cm}$(E-E_{A})/J =
2.8902166871$ \\
$I_3^{(1)}=-\frac{3}{2}$ & $u=0$ & $v=-1$ \\
$I_4^{(1)}=\frac{3}{2}$ & $z_3=-1.1276505247$ & $z_4=+1.1276505247$
\end{tabular}
\end{table}

Note that the pattern of $\{I_3,\ldots,I_{N/2}\}$ is akin to the pattern of
$\{I_1,\ldots,I_{N/2}\}$ for the triplets of a shorter chain. As $N$ grows
larger, the effect of the complex solutions $z_1=z_2^{\ast}$ on the energy
(\ref{eq:e}) relative to that of the real solutions $z_3,\ldots,z_{N/2}$
diminishes and disappears in the limit $N\to\infty$. The two-spinon
singlets will then form a continuum with the same boundaries
(\ref{eq:2spb}).\cite{Woyn82b} Yet the effect of the complex
solutions in the two-spinon singlets will remain strong for
quantities inferred from the Bethe ansatz wave function (for
example, selection rules and transition rates).

How do we solve Eqs.~(\ref{eq:bae-cc-gen})? At $q=\pi$ the real roots
$z_3,\ldots,z_{N/2}$ of the two-spinon singlets are symmetrically
distributed about $z=0$. This observation paves the way for finding
the complex roots exactly. Equation (\ref{eq:bae-cc-3b}) admits a
solution in the limit $|u|\to 0^+, |v|\to 1^+$ along a path with
$|u| \propto (|v|-1)^{1/N}$. This limit matches the divergence on
the left with that in the first term on the right. The
$z_i$ in the non-singular terms on the right have no effect on this solution.

When we substitute the complex roots just found into (\ref{eq:bae-cc-2}), the
left-hand-side becomes $\pm N\pi$, while the sum on the right disappears
because
of the symmetric $z_i$-configuration. Equation~(\ref{eq:bae-cc-2})
thus requires $I^{(2)}=0~({\mathrm mod}~2\pi)$. Finally,
Eqs.~(\ref{eq:bae-cc-1}) with $u=0,\; |v|=1$ can be solved
iteratively similar to (\ref{eq:bae-it-n}), Rapidly converging
solutions $z_3,\ldots,z_{N/2}$ are obtained for the configurations
$I_3,\ldots,I_{N/2}$ indicated in Fig.~\ref{fig:2sps}.

Significant computational challenges arise in the determination of
two-spinon singlet states at $q\neq\pi$, where the real roots are
no longer symmetrically distributed and the complex roots have
$u>0, v>1$. Now there is no way around solving
Eqs.~(\ref{eq:bae-cc-gen}) simultaneously. For this task a good
root-finding algorithm\cite{note2} combined with carefully chosen
starting values will be needed.

The two-spinon triplets play an important role in the
zero-temperature spin dynamics of quasi-1D antiferromagnetic
compounds. They are the elementary excitations of (\ref{eq:H}) which
can be directly probed via inelastic neutron scattering. The
two-spinon singlets, in contrast, cannot be excited directly from
$|A\rangle$ by neutrons because of selection rules. The singlet excitations are
important nevertheless, but in a different context.

Several of the quasi-1D antiferromagnetic compounds are susceptible to a
spin-Peierls transition, which involves a lattice distortion accompanied by an
exchange dimerization. The operator which probes the dimer fluctuations in the
ground state of (\ref{eq:H}) couples primarily to the two-spinon
singlets and not at all to the two-spinon triplets. In a forthcoming
column of this series, the focus will be on transition rates
between the ground state of (\ref{eq:H}) in zero and nonzero
magnetic field and several classes of excited states that are
important for one reason or another.

\section{Suggested problems for further study}

\begin{enumerate}

\item\label{prb:1} Show that $|{\mathcal N}_\pm\rangle$ is not an eigenvector
  of $H_A$. Calculate $\langle {\mathcal N}_\pm|S_n^\alpha
  S_{n+1}^\alpha|{\mathcal N}_{\pm}\rangle, \alpha=x,y,x$, to obtain the
  energy expectation value $\langle {\mathcal N}_\pm|H_A|{\mathcal
  N}_\pm\rangle = -JN/4$ for the N\'eel state. Compare this result with the
  asymptotic ground state energy $\langle A|H_A|A\rangle = (\frac{1}{4} - \ln
  2)JN$ inferred from (\ref{eq:EA}).

\item\label{prb:2} Identify $|A\rangle$ for $N=4,6$ in Tables II and IV of
Part I. Show that the state for $N=8$ with Bethe quantum numbers
(\ref{eq:lambda}) and with real momenta $k_1=1.522002, k_2=2.634831,
k_3=3.64836, k_4=4.761182$ is a solution of (I33) and (I35). Then determine
its energy $E-E_F$ and wave number $k$.

\item\label{prb:3} (a) Use the derivation of (\ref{eq:bae}) from
(I33) and (I35) to establish the relation between the $\lambda_i$
and the $I_i$. Given that all
$\lambda_i$ are integers, show that all $I_i$ are integers for odd
$r$ and half-integers for even $r$. For the given (real) solution
$\{z_1<\cdots<z_{N/2}\}$ determine the one-to-one correspondence
between the elements of (\ref{eq:lambda}) and (\ref{eq:bqn}). (b)
Infer (\ref{eq:e}) from (I30) with $J$ replaced by $-J$ and
(\ref{eq:k}) from (I36).

\item\label{prb:4} Calculate the $\{z_i\}$ of the state $|A\rangle$ iteratively
via (\ref{eq:bae-it-n}) from the $\{I_i\}$ given in (\ref{eq:bqn}) for various
$N$. Compare the results for $N=8$ to the $\{k_i\}$ given in
Problem~\ref{prb:2}. Check the energies for larger systems against the values
listed in Table~I. Extrapolate your finite-$N$ data for $(E_A-E_F)/JN$ and
compare your best prediction and its uncertainty with the exact result
(\ref{eq:EA}).

\item\label{prb:5} (a) Use (\ref{eq:zjk}) and (\ref{eq:sigma-z}) to calculate
the distribution of the magnon momenta in the ground state $|A\rangle$ via
$\rho_0(k)=\int_{-\infty}^\infty dz\, \sigma_0(z) \delta(k-\pi-\phi(z))$. (b)
Express the energy per site of $|A\rangle$ as an integral of the magnon energy,
$-J(1-\cos k)$, weighted by $\rho_0(k)$, and evaluate it to confirm the result
(\ref{eq:EA}).

\item\label{prb:6} In Problem \ref{prb:5} we interpreted $|A\rangle$ as a
  composite of $N/2$ magnons with momentum distribution (\ref{eq:rhok}).
  Increasing $m_z$ from zero means reducing the number of magnons in the ground
  state to $N(\frac{1}{2}-m_z)$. From the solution $\{z_i\}$, calculate the
  finite-$N$ distribution $\rho_N(k_i) = [N(k_i-k_{i+1})]^{-1}$ of magnon
  momenta for $m_z = 0, 0.125, 0.25, 0.375$. Use data for $N=32$ and for a much
  larger system, for example, $N=2048$. Show that for $h=0$, $\rho_N(k_i)$
  converges very rapidly toward (\ref{eq:rhok}).  Show that increasing $h$ does
  not only deplete magnons, it also constricts the range of the magnon momenta.
  
\item\label{prb:7} Generate data for $E(S_T^z)$ via (\ref{eq:bae-it-n}).
  Calculate $h(m_z)$ from (\ref{eq:hmz}) and $\chi_{zz}(m_z)$ from
  (\ref{eq:chizzE}). Then plot $\chi_{zz}$ versus $h$ for $0<h<0.25$ in relation
  to the exact value $1/(\pi^2J)$ at $h=0$. Show that the slope of
  $\chi_{zz}(h)$ increases without bounds as $h\to 0$. Determine the $\chi_{zz}$
  at $h=0$ via extrapolation of the $h\neq 0$ data points, and compare your
  prediction with the exact result (\ref{eq:chi0}).

\item\label{prb:8} The leading singularities in $m_z(h)$ and $\chi_{zz}(h)$ at
  $h=h_S$ can be derived from the $N$-dependence of the three highest energy
  levels in Fig.~\ref{fig:mag}(a): $E(S_T^z)/J - N/4 = 0, -2,
  -4\cos^2(\pi/2(N-1))$, for $S_T^z = N/2, N/2-1, N/2-2$, respectively. To
  calculate the third result, show first that $z_2 = -z_1$ holds. Then derive
  the solution $z_i=\pm\tan(\pi/2(N-1))$. Next use the relations $h_S =
  E(N/2)-E(N/2-1)$, $h = E(N/2-1)-E(N/2-2)$, and $m_z = \frac{1}{2} - 1/N$ to
  express $m_z$ as a function of $h_S-h$. The derivative of the result yields
  (\ref{eq:chi-sat}).

\item\label{prb:9} From Part I we know that the vector with $S_T^z=0$ of a
multiplet is related to the vector with $S_T^z=1$ by an extra magnon with
momentum $k_r=0$ and an extra $\lambda_r=0$, while the $\{k_1,\ldots,k_{r-1}\}$
and $\{\lambda_1,\ldots,\lambda_{r-1}\}$ remain the same. According to
(\ref{eq:zjk}), this result implies $z_r=\pm\infty$. Use (\ref{eq:bae}) to show
that if the $\{z_1,\ldots,z_{r-1}\}$ are to remain the same for the two states,
then the $I_i$-configuration must change as indicated in rows two and three of
Fig.~\ref{fig:Ii}.

\item\label{prb:10} Start from Eqs.~(\ref{eq:bae}) with
$z_1=z_2^*=u+iv$ and real $z_3,\ldots,z_{N/2}$. Use $\phi(x\pm y) =
\arctan(2x/(1-x^2-y^2)) + i\,\atanh(2y/(1+x^2+y^2)) + 2l\pi$ with
(undetermined) integer $l$, and employ the addition theorem
$\phi(z_1) \pm \phi(z_2) = \phi\left([z_1\pm z_2]/[1\pm
z_1z_2]\right)$ in sums and differences to separate real and
imaginary parts of the resulting complex Bethe ansatz equations.
Take into account that $I_1+I_2$ is always an integer.
Adding $N/2$ to $I^{(2)}$ in (\ref{eq:bae-cc-2}), which is
permitted because of the undetermined $2l\pi$'s, makes $I^{(2)}$
assume integer values near zero for the class of states of interest
here.

\end{enumerate}

\acknowledgments
We thank K. Fabricius and A. Fledderjohann for sharing some of their finite-$N$
data with us, and the editors, H. Gould and J. Tobochnik, for their helpful
suggestions.



\begin{thebibliography}{100}
\bibitem{note1} References to results in Part I will be identified by equation
numbers (I\#).
\end{thebibliography}
\end{document}